\documentclass[aps,twocolumn,showpacs]{revtex4}
\usepackage{graphicx}

\begin{document}

\title{Multi-band superconductivity and large anisotropy in FeS crystals}

\author{Hai Lin, Yufeng Li, Qiang Deng, Jie Xing, Jianzhong Liu, Xiyu Zhu$^{*}$, Huan Yang and Hai-Hu Wen}\email{zhuxiyu@nju.edu.cn, hhwen@nju.edu.cn}

\affiliation{Center for Superconducting Physics and Materials,
National Laboratory of Solid State Microstructures and Department
of Physics, National Center of Microstructures and Quantum
Manipulation, Nanjing University, Nanjing 210093, China}

\date{\today}

\begin{abstract}
By using a hydrothermal method, we have successfully grown
crystals of the newly discovered superconductor FeS, which has an
isostructure of the iron based superconductor FeSe. The
superconductivity appears at about 4.5K, as revealed by both
resistive and magnetization measurements. It is found that the
upper critical field is relatively low, with however an rather
large anisotropy
$\Gamma=[(dH_{c2}^{ab}/dT)/(dH_{c2}^{c}/dT)]_{T_c}\approx5.8$. A
huge magnetoresistivity (290$\%$ at 9T and 10K, ${H}$ $\parallel$
c-axis) together with a non-linear behavior of Hall resistivity
vs. external field are observed. A two-band model is applied to fit the magnetoresistance and non-linear transverse resistivity, yielding the basic parameters of the electron and hole bands.
\end{abstract}

 \maketitle

\section{Introduction}
The discovery of high temperature superconductivity in fluorine
doped LaFeAsO has opened a new era for the research on
unconventional superconductivity\cite{HosonoJACS2008}. As far as
we know, there are at least eight different structures concerning
the FeAs-based and FeSe-based
superconductors\cite{ChuCW,WenHH,RGreene}. Beyond the iron
pnictides, so far, many relatives with similar structures have
been found to exhibit superconductivity. For example,
superconductivity has been discovered in iron
chalcogenides\cite{WuMK}, leading to the great expansion of the
iron-based superconducting families. The highest superconducting
transition temperature, as revealed by the traditional criterion,
namely the Meissner effect and zero resistance stays still at
about 55-57K in the 1111 system
\cite{ChenXHNature,RenZACPL,XuZAEPL,ChengPEPL}. In many iron
chalcogenide superconductors, the doping of sulfur leads to the
localization effect of electrons and the system exhibits
insulating behavior in low temperature
region\cite{Petrovic,Petrovic2}. This is probably due to the
narrow bandwidth of the p-electrons in sulfur element. It is thus
a surprise to observe superconductivity at 4.5K in the FeS
system\cite{HuangFQ1} with the typical $\beta$-PbO
structure\cite{WuMK}. A preliminary band structure calculation has
indeed told that the electronic structure as well as the Fermi
surfaces are quite similar in FeSe and FeS systems, namely with
almost identical contributions from the electron and hole
pockets\cite{FeSband}. Thus, it is very curious to know whether
the newly discovered superconductivity in the FeS system has some
similarities as that of the FeSe systems. In this paper,
we report the characterizations of superconductivity in crystals
of FeS. Our results reveal multi-band superconductivity, a rather
large anisotropy of superconductivity, a huge magnetoresistivity
and non-linear Hall effect.

\begin{figure}
\includegraphics[width=8.5cm]{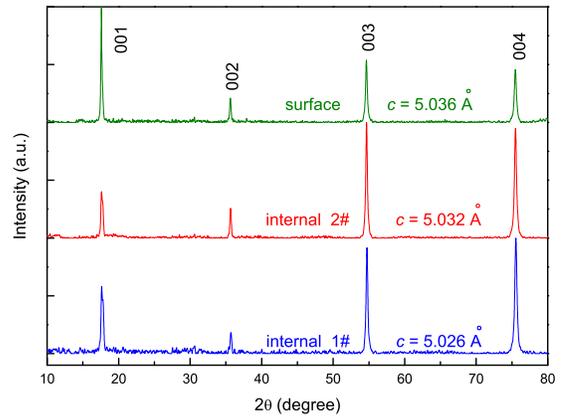}
\caption{(Color online) X-ray diffraction patterns for three
different parts of the same FeS crystal. The one marked with
"surface" is measured on the outside surface of a crystal. Two
curves marked with "internal" are measured on the two inside
surfaces after two times cleaving in the same crystal. }
\label{fig1}
\end{figure}

\section{Experimental details}

In this paper, we report the successful growth of the FeS crystals
using a hydrothermal method\cite{Dong,DuZY}. Firstly,
K${_{0.8}}$Fe${_{1.6}}$S${_2}$ crystals were grown using the
self-flux method. Next, NaOH (J${\&}$K, 99${\%}$ purity) was
dissolved in deionized water in a teflon-linked stainless-steel
autoclave (volume 25 mL). Then, iron powder (Aladdin Industrial,
99.99${\%}$ purity), thiourea (J${\&}$K, 99.9${\%}$ purity), and
several pieces of K${_{0.8}}$Fe${_{1.6}}$S${_2}$ crystals were
added to the solution. After that, the autoclave was sealed and
heated up to 120 $^{\circ}$C followed by staying for 25 hours.
Finally, all the potassium atoms are extracted from the body and
the FeS crystals can be obtained by leaching.

In this study, X-ray diffraction (XRD) measurements are performed
on a Bruker D8 Advanced diffractometer with the Cu-K$_\alpha$
radiation. DC magnetization measurements are carried out with a
SQUID-VSM-7T (Quantum Design). The resistive measurements are
done with the four-probe method on a Quantum Design instrument
Physical Property Measurement System (PPMS). And the Hall
resistive measurements are done with the six-probe method.

\begin{figure}
\includegraphics[width=8.2cm]{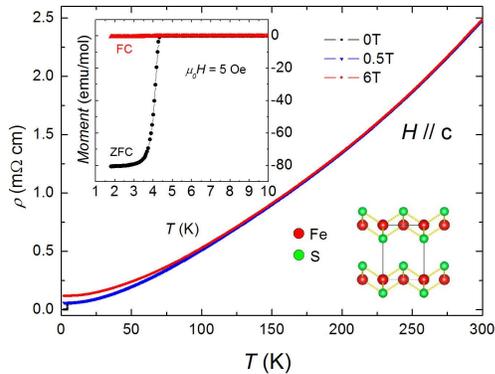}
\caption{(Color online) Temperature dependence of resistivity for
the FeS crystal at magnetic fields of 0, 0.5 and 6 T. The upper left inset is the
temperature dependence of magnetic susceptibility measured in both
ZFC and FC modes, with an applied field of 5 Oe parallel to
c-axis. The lower right inset shows a schematic structure of
tetragonal FeS.} \label{fig2}
\end{figure}

\section{Results and discussion}
\subsection{Sample characterization}

In Fig.~\ref{fig1}, we show the X-ray diffraction (XRD) spectra
for three different parts of an FeS crystal. The sample has a very layered feature and can be easily cleaved. One XRD
pattern is taken from the external surface of the crystal directly
and two measurements are done on the inner surfaces after
cleaving the same crystal. One can see that only (00l) reflections
can be seen in all the spectra, yielding a c-axis lattice constant
$c=5.0310\pm0.0050\AA$. Both inside and outside parts show similar
behavior and the ${c}$ lattice constants are close to
$c=5.0307\AA$ in the previous report\cite{HuangFQ1}, indicating
a uniform bulk property of our FeS crystal with a tetragonal
structure.

The temperature dependence of resistivity at various magnetic
fields are shown in Fig.~\ref{fig2}. The lower right inset of Fig.2 shows the
schematic crystal structure of tetragonal FeS. With the lowering down of temperature, the resistivity decreases monotonically, which shows a highly metallic conductivity. The residual resistivity ratio, defined as
RRR=$\rho(300K)/\rho(0K)\approx $46, is quite large, indicating
the good quality of the sample. It is interesting to have a comparison with the FeSe single crystals\cite{WuMK,Hurongwei,Matsuda}. In the early FeSe samples, the RRR in FeSe was reported less than 15\cite{WuMK,Hurongwei}. However, a newly updated value of RRR=$\rho(300K)/\rho(0K)\approx $80 can be estimated in a clean FeSe single crystal\cite{Matsuda}. Having a glance at the temperature dependence of resistivity in FeSe and FeS, one can immediately find the difference. The temperature dependent resistivity in FeS exhibits a positive curvature all the way up to 300 K, not like
that in FeSe where a negative curvature is generally observed in the high temperature region\cite{WuMK,Hurongwei,Matsuda}. With the increase of applied magnetic field, superconductivity is quickly suppressed (${H}$ $\parallel$ c-axis), and a significant magnetoresistance (MR) is also observed. In zero
field, an abrupt resistivity drop can be seen at $T_{c}$ = 4.5K. This superconducting transition can also be seen in the temperature dependence of the magnetic susceptibility measurements with the zero-field-cooled (ZFC) and field-cooled(FC) modes shown in the upper left inset of Fig.2. Since the $M(T)$ curve measured in the ZFC mode is flattening in the low temperature limit, and the XRD data shows the pure FeS phase, we can conclude the perfect superconducting shielding of the sample at 5 Oe.

\subsection{Magnetic and transport properties}

\begin{figure}
\includegraphics[width=8.5cm]{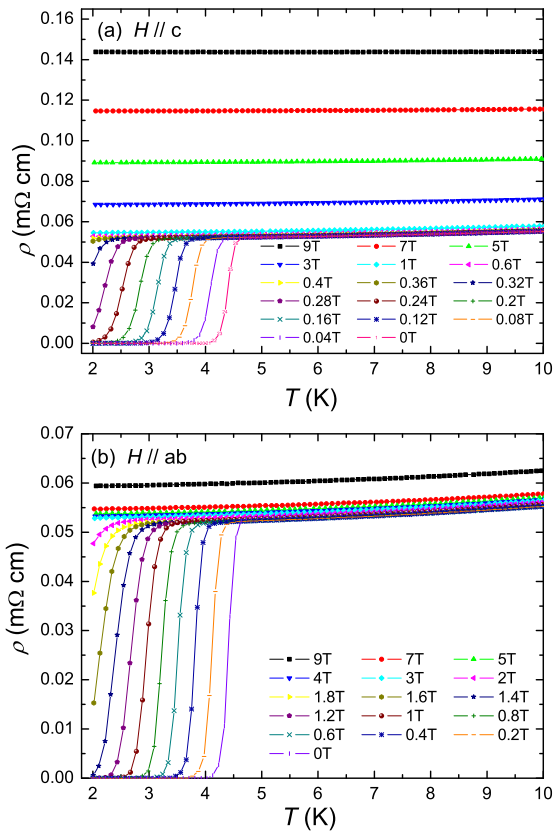}
\caption{(Color online) Temperature dependence of resistivity for
the FeS crystal at zero field and various magnetic fields with the
field directions of (a) ${H}$ $\parallel$ c-axis and (b) ${H}$
$\parallel$ ab-plane.} \label{fig3}
\end{figure}

Fig.~\ref{fig3}(a) and (b) present the resistivity data measured
under different magnetic fields of (a) ${H}$ $\parallel$ c-axis
and (b) ${H}$ $\parallel$ ab-plane, while the current is always
applied in the ab-plane. The superconducting transition appears at
4.5K at an ambient field. As the applied field is increased, the
superconducting transition is suppressed down to 2K gradually. The
superconductivity at 2K vanishes at a magnetic field of only 0.36
T when ${H}$ $\parallel$ c-axis, or about 2T when ${H}$
$\parallel$ ab-plane. This indicates a quite
large anisotropy. Furthermore, the behavior of
magnetoresistivity under high fields for the two directions
are very distinct. The magnetoresistance can reach 180\% at
${T}$=5K with $\mu_0H$=9T when the magnetic field is parallel to
c-axis. For some samples, the MR value can even reach 290\% (see below). This value, as far as we know, is much higher than many iron based materials, including the BaFe$_2$As$_2$ parent
phase\cite{BaFe2As2MR} and the NdFeAsO parent
phase\cite{ChengPPRB}. However, it is comparable with the MR value in the recently reported clean FeSe crystal\cite{Matsuda}. In the two band model, the magnetoresistance can be enhanced when the charge carriers
from the two bands have opposite signs. This explains why the magnetoresistance effect is very strong when the field is applied parallel to the c-axis. Similar case occurs in the MgB$_2$ system with electron and hole contributions\cite{YangHuanPRL2008}. When the field is applied
along the ab-plane, as shown in Fig.3(b), the magnetoresistance is only about 13\% at 9T
and 5K. For different samples, this value can vary a little bit, but is still very small. This strongly indicates a two-dimensional feature of the
electric conduction.

\begin{figure}
\includegraphics[width=8.5cm]{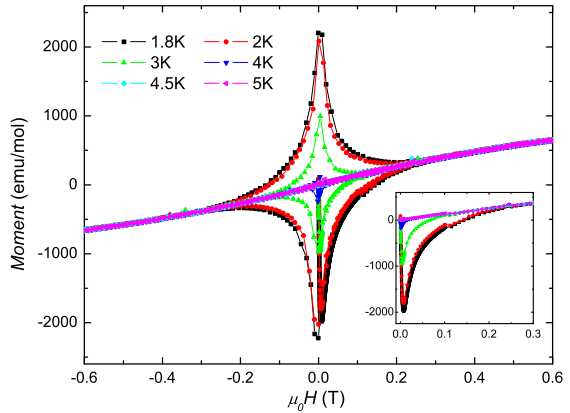}
\caption{(Color online) Magnetization hysteresis loops of the FeS
crystal at various temperatures below $T_c$. The inset shows an
enlarged view of the MHLs in the magnetic field penetration
process. The full magnetic penetration field at 1.8 K is only about 85 Oe.} \label{fig4}
\end{figure}

Fig.~\ref{fig4} presents the magnetization hysteresis loops (MHLs)
of the FeS crystal at various temperatures with the magnetic field applied parallel to c-axis. A weak and soft ferromagnetic
background can be seen above $T_c$. No hysteresis of this
ferromagnetic background is observed here. This weak ferromagnetic
signal may come from the magnetic impurities or it is an intrinsic feature, which
needs to be further resolved in the future. From the MHLs, we also
determined the width of the magnetization in the field
ascending and decreasing processes. The shape of MHLs and the
hysteresis indicate that FeS is a type-II superconductor. The inset shows an enlarged view of the magnetization in the magnetic penetrating process. One can see that the full penetrating field which corresponds to the maximum value of magnetization is only about 85Oe (1.8K). This suggests either a low charge carrier density, or an easy vortex motion with weak flux pinning effect.

\begin{figure}
\includegraphics[width=8.5cm]{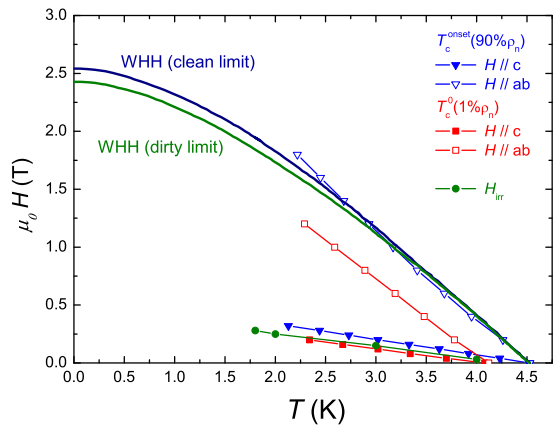}
\caption{(Color online) The phase diagram of the FeS crystal,
where the upper critical fields are determined by using three
different criterions. The filled symbols represent the upper
critical fields ${H_{c2}^c}$ under ${H}$ $\parallel$ c-axis determined with the data shown
in Fig.~\ref{fig3}(a), while the hollow ones represent
${H_{c2}^{ab}}$ under ${H}$ $\parallel$ ab-plane determined with the data shown in
Fig.~\ref{fig3}(b). The filled green circles represent ${H_{irr}}$
determined from the irreversible magnetization shown in
Fig.~\ref{fig4}. The solid lines are the WHH fitting results in
clean and dirty limit.} \label{fig5}
\end{figure}

In Figure 5, we show the upper critical
field $H_{c2}$ and irreversible field $H_{irr}$ versus temperature. To determine the upper critical field ${H_{c2}}$, we can linearly extrapolate the resistivity data between 5K and 10K down to low temperature region as the normal-state value $\rho_{n}$, and
then determine $T_c^{onset}$ with the criterion of 90\% $\rho_{n}$, and $T_c^0$ with 1\% $\rho_{n}$, respectively. The irreversibility line
${H_{irr}}$(T) is determined from the irreversible magnetization
shown in Fig.~\ref{fig4} by using the criterion of $\Delta M$ = 20
emu/mol. This criterion is chosen since our instrument gives an opening width of magnetization, i.e.,  $\Delta M$ = 20 emu/mol in the normal state (5K). For a FeS sample with 1 mg mass, this corresponds to the criterion of 2.27$\times 10^{-4}emu$. The reason for this may be the slight diamagnetic signal coming from the organic Teflon tape, or induced by the eddy current of the supporting copper tube.
The upper critical fields curves $H_{c2}$(T) look rather straight and even slightly positively curved. This may be induced by the multi-band effect\cite{MultibandHc2}. We also add the theoretical curves for $H_{c2}$(T) of the Werthamer-Helfand-Hohenberg (WHH) theory in the dirty and clean limit. One can see that, in the temperature region of 2.5K to $T_c$, the experimental data are close to the theoretical curves. Because of the linearity of $H_{c2}$(T) near $T_c$, the data can be easily fitted with a linear line with the slopes
${[d\mu_0H_{c2}^c/dT]_{onset}}$=-0.13392(T/K),
${[d\mu_0H_{c2}^{ab}/dT]_{onset}}$=-0.77861(T/K) with the magnetic fields parallel to c-axis and ab-planes, respectively. Using the data with the criterion of 90\% $\rho_{n}$, we can get the upper critical field $H_{c2}$ by using the WHH formula\cite{WHH formula} in the dirty limit
${H_{c2}=-0.69T_{c}[dH_{c2}/dT]_{T_c}}$, which gives
${\mu_0H_{c2}^c}$(0)=0.42T and ${\mu_0H_{c2}^{ab}}$(0)=2.4T. The
anisotropy $\Gamma$ determined by the ratio of the upper critical
field along the two different directions is about 5.8. This value
is much higher than most 11, 111 and 122 iron-based
superconductors\cite{Yuan,M.A.Tanatar,LiChunhongPRB}, but
comparable to that in 1111 system\cite{JiaYinAPL}. We became aware of a recent report on the FeS crystal synthesized in the similar way as ours\cite{Paglione}, the authors report an anisotropy of about 10 for the FeS system.

\begin{figure}
\includegraphics[width=8.5cm]{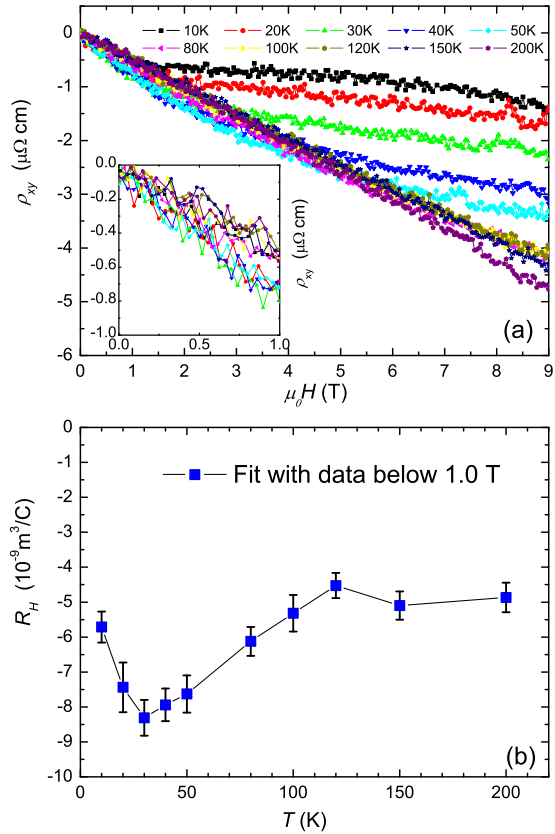}
\caption{(Color online) (a) Magnetic field dependence of the
transverse resistivity $\rho_{xy}$ at different temperatures. (b)
The temperature dependence of Hall coefficient $R_H$ determined
using the data of $\rho_{xy}$ vs. ${H}$ in the field region
below 1T. The error bar is derived from the scattering range of $\rho_{xy}$ data.} \label{fig6}
\end{figure}

The Hall resistance and the magnetoresistance measurements shown below are
done on a separate sample. In Fig.~\ref{fig3}(a), we have showed a
very strong MR effect observed with $H$ $\parallel$ c-axis. A
straightforward understanding to the large MR would be
the multi-band effect. From the band structure
calculations\cite{FeSband}, indeed, both the hole and electron
pockets appear in identical Fermi surface area. We thus have measured
the Hall resistivity $\rho_{xy}$ and present the data at various
temperatures in Fig.~\ref{fig6}(a). For temperatures below 80K,
$\rho_{xy}$ does not have a linear correlation with the magnetic field. While, interestingly, the Hall resistivity $\rho_{xy}$ exhibits a kinky feature at temperatures of 10K and 20K. Below a certain field, the
$\rho_{xy}$ shows a rough linear behavior. We must emphasize that
for a simple two-band system, the Hall resistivity $\rho_{xy}$ vs.
$H$ may be non-linear, but normally it will not show a kinky
feature. We would assume that this kinky structure may be observed by accident in the low temperature region due to the multi-band effect. However, with increasing temperature, the
non-linear $\rho_{xy}$ curves at low temperatures seem to
evolve gradually to the linear ones at high temperature (above 80K). So we can
assume that at high temperatures, one main band makes most
contribution to the conduction. As the temperature is cooled down,
multi-band effect emerges. We thus calculate the Hall
coefficient $R_H$ using the low-field part of each curve below
1T and show the results in Fig.~\ref{fig6}(b). It is very interesting that
the Hall coefficient ${R_H}$ is generally negative showing a
dominance of the electron-like charge carriers. In the low temperature region, there is a non-monotonic temperature dependence of $R_H$ vs. $T$. This behavior is different from that in many
FeAs-based systems in which the Hall coefficient, no matter
positive or negative in sign, increases the magnitude in lowering
the temperature\cite{ChengPPRB,BaFe2As2Hall,SrFeAsFHanFei}.
This enriched message from the Hall effect measurements must
reflect an interesting multi-band effect.

\subsection{Multi-band analysis on the transport data}

\begin{figure}
\includegraphics[width=8.5cm]{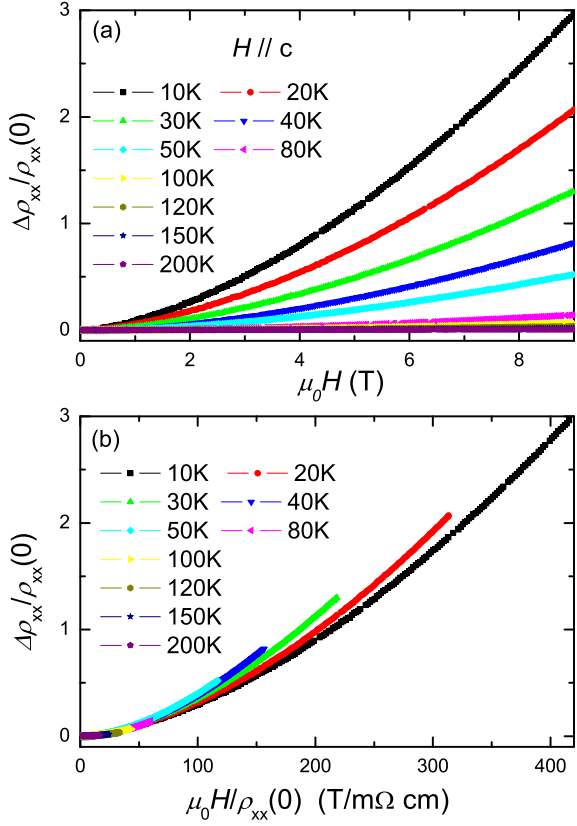}
\caption{(Color online) (a) Temperature dependent magnetoresistance measured with $H\parallel c$-axis. A huge MR effect is observed at 9 T and 10 K. (b) The Kohler plot at different temperatures
of the sample, and the Kohler's rule is obviously violated.} \label{fig7}
\end{figure}

In Fig.~\ref{fig7}(a), we present the MR effect when the current
is applied parallel to the ab-plane and the magnetic field is aligned
along $c$-axis. One can see that the MR can reach about 290\%
at 9 T and 10 K, and this value is rather high, even considering those in the parent
phase with the presence of the antiferromagnetic
order\cite{ChengPPRB,BaFe2As2Hall}. We have also tried the Kohler's plot scaling of the
magnetoresistance and the scaling is shown in Fig.~\ref{fig7}(b). One can find an
obvious violation of the Kohler's scaling rule, which can be attributed
to the multi-band effect in this system.

The huge MR effect together with the violation of Kohler's rule can certainly get an explanation from the multi-band effect, since the
multiple scattering rate $\tau_i$ (or mobility $\mu_i=e\tau_i/m_i$) and the charge carrier density $n_i$ ($i=$ band
index, and $m_i$ is the effective mass) entangle each other and contribute a large MR effect when
the scattering rate of each band has different temperature
dependence\cite{YangHuanPRL2008}. In the system with the isotropic mobility and effective mass for each band, the longitudinal and transverse conductance tensor components can be expressed as
\begin{equation}
\sigma_{xx}(B)=\frac{\rho_{xx}(B)}{\rho_{xx}^2(B)+\rho_{xy}^2(B)}=\sum_i^q \frac{\sigma_i}{1+{\mu_i}^2B^2},\label{eqxx}
\end{equation}
\begin{equation}
\sigma_{xy}(B)=\frac{\rho_{xy}(B)}{\rho_{xx}^2(B)+\rho_{xy}^2(B)}=\sum_i^q \frac{\sigma_i\mu_iB}{1+{\mu_i}^2B^2}.\label{eqxy}
\end{equation}
Eqs.~\ref{eqxx} and \ref{eqxy} are used to describe the transport properties of the system with $q$ types of charge carrier, and $\sigma_i=n_ie^2\tau_i/m_i$ is the conductance for the $i^\mathrm{th}$ band at zero magnetic field. However, neither the two-band model ($q=2$) nor the three-band one ($q=3$) can fit the experimental data very well by using two or three sets of $\sigma_i$ and $\mu_i$ (results not shown here), which may suggest that the Fermi surface for each band is more complex than a simple cylindric shaped one, or the effective mass or the mobility is not a constant for each band. In a two-band model in the low field region, the conductances can be expressed as the polynomial expansion form approximately to the third terms,
\begin{equation}
\sigma_{xx}(B)\approx\sum_{i=1}^2\sigma_i-\left(\sum_{i=1}^2{\sigma_i}{\mu_i}^2\right)B^2+\left(\sum_{i=1}^2{\sigma_i}{\mu_i}^4\right)B^4,\label{eqxx2}
\end{equation}
\begin{equation}
\sigma_{xy}(B)\approx \sum_{i=1}^2\sigma_i\mu_iB-\left(\sum_{i=1}^2{\sigma_i}{\mu_i}^3\right)B^3+\left(\sum_{i=1}^2{\sigma_i}{\mu_i}^5\right)B^5.\label{eqxy2}
\end{equation}

\begin{figure}
\includegraphics[width=8.5cm]{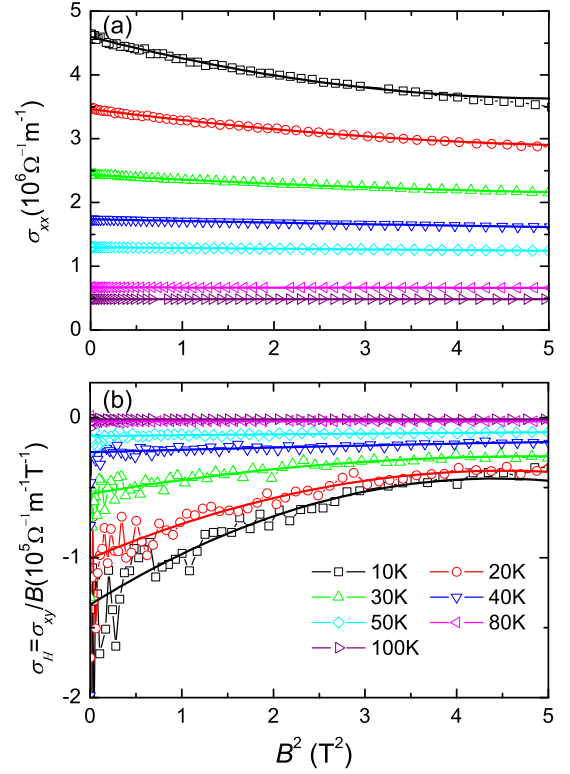}
\caption{(Color online) $B^2$ dependent (a) longitudinal and (b) Hall conductivities and two-band fit in the small-field range. The fitting range of the magnetic field is 0 - 2 T for the data measured at 10 K, 20 K and 30 K, 0 - 3 T for the data measured at 40 K and 50 K, and 0 - 4 T for the data measured at 80 K and 100 K.} \label{fig8}
\end{figure}
\begin{figure}
\includegraphics[width=9cm]{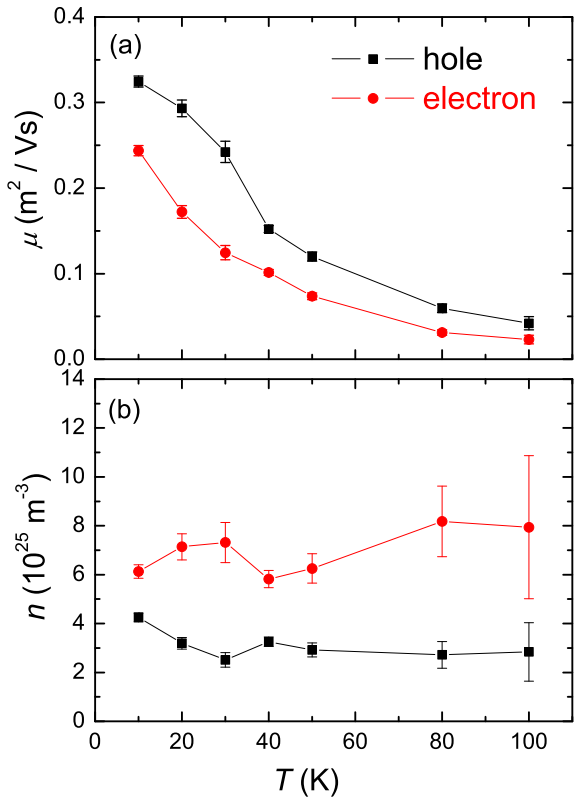}
\caption{(Color online) Temperature dependence of the (a) mobility and (b) charge carrier density for electron and hole bands derived from the two-band model fitting. The error bars here are determined from the free fitting process.} \label{fig9}
\end{figure}

In the case of approximate two-band model, we have only four parameters, namely $\mu_i$ and $n_i$ (i=1, 2) that need to be resolved. According to above equations, at each temperature, we have two curves $\sigma_{xx}$(B) and $\sigma_{xy}$(B) to be fitted in order to yield the wanted parameters. We use Eqs.~\ref{eqxx2} and \ref{eqxy2} to fit the experimental data in the low field region, and the fitting results are shown in Fig.~\ref{fig8} with the solid lines. For different temperatures, the fitting ranges of magnetic field are slightly different. The fitting range is 0 - 2 T for the data measured at 10 K, 20 K and 30 K, 0 - 3 T for 40 K and 50 K, and 0 - 4 T for 80 K and 100 K. The parameters resulting from the fitting are given in Fig.~\ref{fig9}. It should be noted that only the scattering rate $\tau_i$ is expected to have the temperature dependence, and $n_i$ for each band should be constant at different temperatures. As shown in Fig.~\ref{fig9}(b), $n_i$ for both bands have indeed very weak temperature dependence, which makes the fitting more reliable. From the results derived here, we can find that there should be both an electron band and a hole band in this material, and the two kinds of charge carriers have almost balanced mobilities and charge carrier densities at different temperatures. This is consistent with the theoretical results of band structure calculations that the electron and hole like charge carriers contribute almost identically in FeS\cite{FeSband}. Although we have concluded only multi-band electric conduction from the normal state properties, the multi-band superconductivity is naturally imaginable, since superconductivity is evolved from the normal state. Recent thermal conductivity and specific heat data give strong support to the multi-band superconductivity\cite{LiSY,FeSSH}. Our detailed characterization of superconductivity and
the normal state properties, especially the strong
magnetoresistance and high anisotropy in FeS will stimulate further investigations in this new superconductor, and help to reach the final
understanding to the mechanism of the iron-based superconductors.

\section{Conclusions}
In summary, we have successfully grown crystals of FeS with a
tetragonal structure. Resistive measurements reveal that the
anisotropy determined from the slopes of upper critical fields
near $T_c$ under two configurations ${H}$ $\parallel$ ab-plane and
${H}$ $\parallel$ c-axis is about 5.8. Further resistive
measurements reveal a very strong magnetoresistance (up to 290$\%$ at 9T and 10K) when ${H}$
$\parallel$ c-axis and a clear non-linear Hall effect. Detailed analysis based on the approximate two-band model give rise to the basic parameters of the electron and hole bands, showing almost balanced contributions of these two bands. All these suggest
the importance of multi-band effect and rather high anisotropy of the new superconducting FeS system.

\section*{ACKNOWLEDGMENTS}
This work was supported by the National Natural Science Foundation
of China (Grant No. 11534005, No.11190023), the Ministry of Science and
Technology of China (Grant Nos.2011CBA00102, 2012CB821403).

\end{document}